\documentclass[twocolumn,aps,prb,longbibliography,superscriptaddress,floatfix]{revtex4-2}
\usepackage{nicefrac}
\usepackage{amsfonts}
\usepackage{mathtools,amssymb,graphicx,units,bm}
\usepackage{amsmath}
\usepackage{subfigure}
\usepackage{multirow} 
\usepackage{tabularx} 
\usepackage{array}

\usepackage[plainpages=false,pdfpagelabels,colorlinks=true,linkcolor=red,urlcolor=PineGreen,citecolor=PineGreen,pdftitle={Title},pdfauthor={},pdfdisplaydoctitle=true,pdfduplex=DuplexFlipLongEdge]{hyperref}

\usepackage{soul} 
\usepackage[dvipsnames,usenames]{color}
\usepackage[normalem]{ulem}
\graphicspath{{figures/}}
\usepackage{color}
\usepackage{bbold} 
\usepackage{float}
\usepackage{silence}
\usepackage{dcolumn}
\usepackage{physics}%
\usepackage[dvipsnames]{xcolor}
\emergencystretch=\maxdimen
\usepackage{units}
\usepackage{tensor} 
\usepackage{braket}
\usepackage{hyperref}
\usepackage{amsfonts}
\usepackage{fontenc}
\usepackage{graphicx}
\usepackage{xcolor}
\usepackage{textcomp}
\usepackage{epstopdf}
\usepackage{braket}
\usepackage{mathtools}
\usepackage{amsmath}
\usepackage{dcolumn}
\usepackage{multirow}
\usepackage{units}
\usepackage{ulem}
\usepackage{mathrsfs, amssymb, bm}

\begin{document}

\title{Orbital magnetization in Sierpinski fractals}

\author{L. L. Lage}
\email{lucaslage@id.uff.br}
\affiliation{Instituto de F\'isica, Universidade Federal Fluminense, Niter\'oi, Av. Litor\^{a}nea sn 24210-340, RJ-Brazil}

\author{Tarik P. Cysne}
\email{tarik.cysne@gmail.com}
\affiliation{Instituto de F\'isica, Universidade Federal Fluminense, Niter\'oi, Av. Litor\^{a}nea sn 24210-340, RJ-Brazil}

\author{A. Latg\'e}
\affiliation{Instituto de F\'isica, Universidade Federal Fluminense, Niter\'oi, Av. Litor\^{a}nea sn 24210-340, RJ-Brazil}

\date{\today}

\begin{abstract}
 
 Orbital magnetization (OM) in Sierpinski carpet (SC) and triangle (ST) fractal is theoretically investigated by using Haldane model as a prototypical example. The OM calculation is performed following two distinct approaches; employing the definition and local markers formalism. Both methods coincides for all systems analyzed. For the SC, higher fractal generations create a dense set of edge states, resulting in a staircase profile, leading to fluctuations in the magnetization as a function of the chemical potential. In contrast, the ST self-similarity produces distinct fractal-induced spectral gaps, which manifest as constant plateaus in the magnetization. The STs exhibit a pronounced sensitivity to edge terminations. Our results reveal how quantum confinement in fractal structures affects the electronic orbital angular momentum, pointing to possible pathways for exploring novel orbitronics in systems with complex geometries.
\end{abstract}

\maketitle


\section{Introduction} Fractals are mathematical structures exhibiting self-similarity and non-integer dimensions \cite{Mandelbrot}. Fractal structures are often found in nature and appear in contexts such as molecular chemistry \cite{Sadegh2017, Sendker2024}, biology \cite{Smith1989, Wang2024}, and geography \cite{Jiang2016}. Of the many types of fractals, the Sierpinski fractals are some of the most studied. They are constructed from simple geometric shapes by iteratively removing specific regions, resulting in structures that exhibit self-similarity. For instance, the Sierpinski carpet (SC) is constructed through an iterative process of subdividing a solid square into a $ 3 \times 3 $ grid of smaller congruent squares and removing the central square. This process is repeated recursively for each of the remaining eight squares, resulting in a structure composed of self-replicated solid squares. An analogous algorithm can be used to generate the Sierpinski triangle (ST) fractal.

In physics, electronic fractals have gained increasing attention in recent years. With advances in the manipulation of matter at the nanoscale, the fabrication of nanostructured fractal patterns has become possible. Electronic nanoscale fractal systems have been constructed using various approaches, including bismuth nanostructures \cite{Canyellas2024}, surface-positioned molecules \cite{Kempkes2018}, and self-assembled molecular systems \cite{Newkome2006, Shang2015}, where electrons are constrained to move within fractal lattice geometries. The non-integer fractal dimension of this structure enables the emergence of many interesting and exotic phenomena. A remarkable example is the recent proposal in which electronic fractals can host topological phases even without the usual driving terms, such as spin-orbit coupling or magnetic fields, in the electronic Hamiltonian \cite{CristianePRL}. We also mention the possible loss of correlation between transport coefficients and topological invariants in certain electronic fractals, as pointed out by some numerical studies \cite{Iliasov2020, Fremling2020}. Electronic fractals have also been known to exhibit particular charge localization \cite{lagefrontier, CristianePRR} and self-similarity in electronic responses \cite{Pedersen2020,Agarwala2019,lagepccp,BiplabPRB}. Furthermore, fractal systems lack translational symmetry, making them interesting platforms for studying local topological aspects of matter \cite{Bianco2011, dosAnjos2025, Melo2023, Bau2024}. In fact, real-space topology has been studied in fractal systems of both $\mathbb{Z}_2$ insulators and Chern insulators, including Sierpinski triangles \cite{haldanecristiane, PhysRevB.98.205116,SPaiPRB} and Sierpinski carpets \cite{RojoFrancas2024,lagetarik2025}. Other types of topological phases have also been explored in distinct fractal systems \cite{Manna_Roy_2023,Chen2025,lage2025sc,Cao2025}.

Recently, electronic orbital angular momentum in condensed matter systems has piqued the interest of the scientific community. Its relevance to the equilibrium magnetization of nanostructured systems \cite{Tschirhart2021, Peredkov2011, liu2025} and the development of novel approaches for generating nonequilibrium orbital currents \cite{Cysne2021, Go2018} and densities \cite{Cysne2023, Johansson2024} has given rise to the emerging field of orbitronics \cite{Cysne2025, Jo2024, Burgos2024}.

\begin{figure*}[t!]
    \centering
    \includegraphics[width=0.85\textwidth]{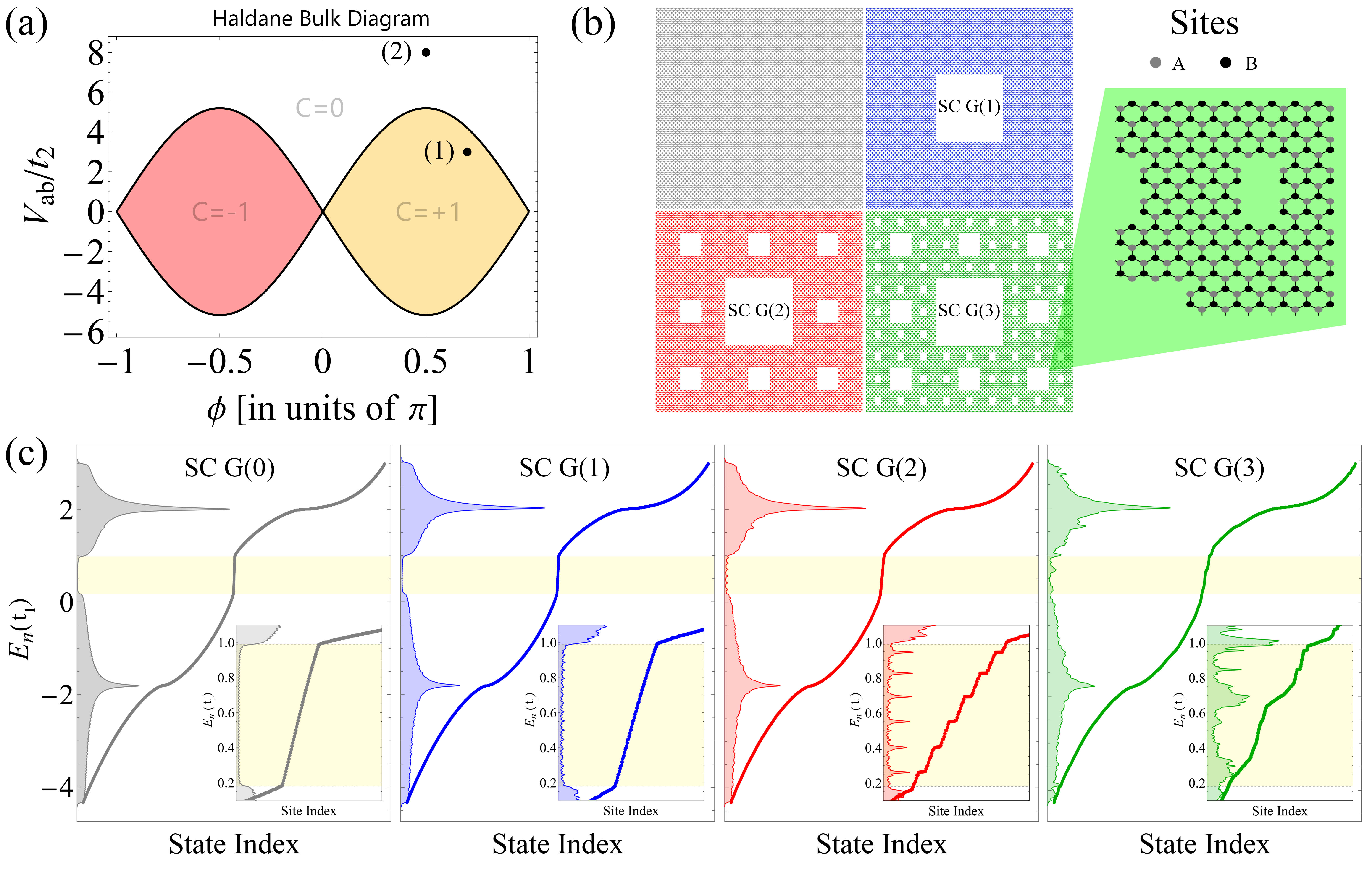} 
    \caption{(a) Topological phase diagram of the 2D Haldane model. \textcolor{black}{The red marks indicate the parameters used in numerical calculations. Parameters (1) and (2) correspond to $V_{ab}/t_2=3.0,8.0$ and $\phi=0.7\pi,0.5\pi$, respectively.}  (b) Four generations of the Sierpinski carpet (SC), with a magnified view of the underlying honeycomb lattice. (c) Energy spectra and density of states for SC G(0-3), \textcolor{black}{considering point (1) in the phase diagram of panel (a).} The inset shows in-gap states from the topological phase. Yellow shading indicates the Haldane bulk energy gap  for parameters, $t_2=t_1/3$, $V_{ab}/t_2=3.0$ and $\phi=0.7\pi$ .}
   \label{FIG1}
\end{figure*}
Understanding how magnetism manifests itself in fractal geometries remains an intriguing open question \cite{Conte2024}, especially in light of advances in nanoscale manipulation and the growing range of materials used in nanoscience. In particular, studying the orbital contribution to the magnetization of fractal nanostructures is appealing, given the strong dependence of the electronic spectrum on the specific fractal geometry. To explore these effects, we study the Haldane model, a paradigmatic example exhibiting orbital magnetization, on two different Sierpinski fractal architectures. Firstly, we present results  for the Sierpinski carpet, focusing on the evolution of equilibrium orbital magnetization of the system, with fractal generation of the Sierpinski carpets. A subsequent analysis for Sierpinski triangles highlights the emergence of fractal-induced energy gaps and the crucial role of edge termination geometry.

\section{Model and Methods} 
In our study, we consider electrons described by the Haldane Hamiltonian, which can be written in real space as,
\begin{eqnarray}
\mathcal{H}_{\rm H}&=&t_1\sum_{\langle {\bf i}, {\bf j} \rangle}c^{\dagger}_{{\bf i}}c_{{\bf j}}+V_{\rm ab}\sum_{\bf i}\tau_{\bf i}c^{\dagger}_{{\bf i}}c_{{\bf i}}\nonumber \\
&&\hspace{0mm}+t_2\sum_{\langle \langle {\bf i}, {\bf j}\rangle \rangle} e^{i \nu_{{\bf i}{\bf j}} \phi} c^{\dagger}_{{\bf i}} c_{{\bf j}} \ \  \text{+  \ H.c.}
\ \ , \label{HHaldane}
\end{eqnarray}
where $t_1$ is the intensity of hopping between nearest-neighbor sites $\langle {\bf i},{\bf j}\rangle$ in the underlying honeycomb structure. $V_{\rm ab}$ is the intensity of sublattice potential with, $\tau_{\bf i}=+1(-1)$ for the sites at sublattice $A(B)$. The strength coupling $t_2$ between the next-nearest-neighbor sites ($\langle \langle {\bf i}, {\bf j}\rangle \rangle$) captures the effect of the staggered flux $\phi$ introduced in the model \cite{Haldane1988}. In this term, $\nu_{{\bf ij}}=+1(-1)$ for clock (counter clock)-wise orientations of hoppings. The illustrative topological phase diagram of this model in the 2D bulk system is shown in Fig.~\ref{FIG1} (a). It is composed of two topologically non-trivial phases with Chern numbers $C=\pm 1$ and a topologically trivial phase with zero Chern number ($C=0$). Unless explicitly mentioned, calculations were performed adopting, $t_2=t_1/3$, $V_{ab}/t_2=3.0$, and $\phi=0.7\pi$.

\begin{figure*}[t!]
    \centering
    \includegraphics[width=0.99\textwidth]{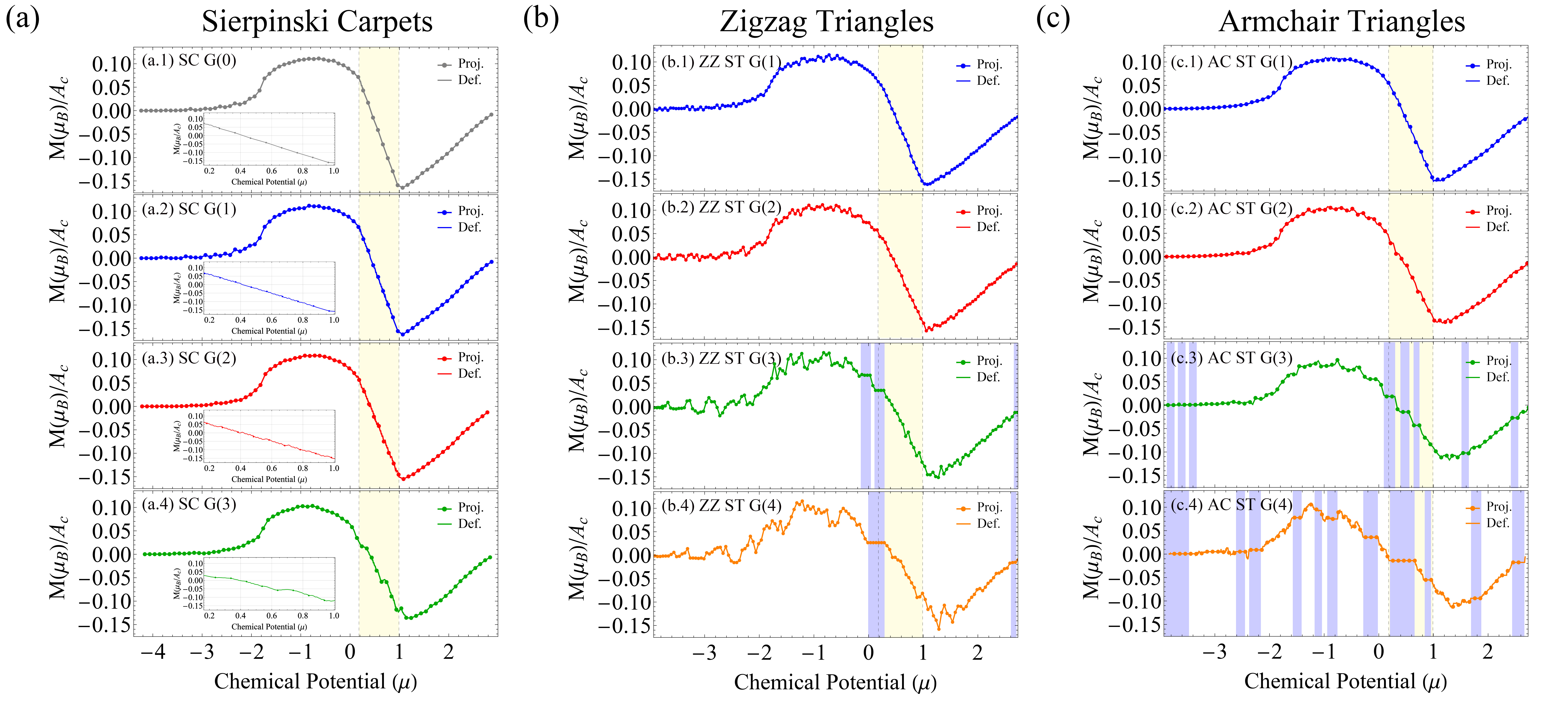} 
    \caption{\textcolor{black}{(a) Orbital magnetization vs. chemical potential in the Haldane model for Sierpinski carpet generations G(0$-$3) (a.1$-$4) and also for Sierpinski triangles with (b) zigzag and (c) armchair termination for generations G(1$-$4) (b.1$-$4) and (c.1$-$4), respectively. Solid lines represent the results obtained using the definition of orbital magnetization [Eq. \ref{Mdef}], while circles represent the results obtained from the local markers formulation [Eqs.(\ref{orbitalMarker}-\ref{Mmarker})]. 
    The shaded yellow region corresponds to 2D bulk gap in the Haldane model with $V_{ab}/t_2=3.0$, $\phi=0.7\pi$. Blue areas in panels (b.3$-$4) and (c.3$-$4) highlight fractal-induced gaps in ZZ and AC STs. The insets in panels (a.1$-$4) show magnified views  of orbital magnetization in the 2D bulk gap region for SC.} }
   \label{FIG2+FIG4}
\end{figure*}
The Haldane model breaks the time-reversal symmetry, enabling equilibrium orbital magnetization. As a simple electronic model, it has served as a paradigm for exploring new aspects of this phenomenon \cite{Bianco2013a, Thonhauser2005, Ceresoli2006}. Here, we consider the Haldane model in fractal geometries and examine the impact of fractality on orbital magnetization.

The magnetization arising from electronic orbital angular momentum can be defined as:
\begin{eqnarray}
M=-\frac{e}{2cA}\sum_{E_n < \mu}\bra{\phi_n}{\bf r}\times {\bf v}\ket{\phi_n}, \label{Mdef}
\end{eqnarray}
where, $A$ is the area of the sample, ${\bf v}=-\frac{i}{\hbar}\left[{\bf r},\mathcal{H} \right]$ is the velocity operator, and $\ket{\phi_n}$ are eigenstates of the real-space Hamiltonian $\mathcal{H}$ with eigenenergies $E_n$. The summation in Eq. (\ref{Mdef}) runs over all the occupied states, $E_n< \mu$, where $\mu$ is the chemical potential. The orbital magnetization can be written as $M = -\frac{ie}{2\hbar c A} \sum_{E_n < \mu} (\langle\phi_n| x \mathcal{H} y |\phi_n\rangle - \langle\phi_n| y \mathcal{H} x |\phi_n\rangle)$.
Bianco and Resta showed in Ref. \cite{Bianco2013a} that Eq. (\ref{Mdef}) can be expressed as a sum over a local quantity in real space. By writing the ground state projector,
$\mathcal{P}=\sum_{E_{n}<\mu}\ket{\phi_n}\bra{\phi_n}$,
and its complement $\mathcal{Q}=\mathbb{1}-\mathcal{P}$, one can define a local mathematical object, i.e., a marker, that plays the role of an orbital magnetic moment density:
\begin{eqnarray}
m({\bf r})&=&m_{\rm LC}({\bf r})+m_{\rm IC}({\bf r})+m_{\rm BC}({\bf r}),  \label{orbitalMarker}\\
m_{\rm LC}({\bf r})&=&\frac{e}{\hbar c}\text{Im}\bra{{\bf r}}\mathcal{P} x \mathcal{Q} \mathcal{H} \mathcal{Q} y \mathcal{P}  \ket{{\bf r}},
\label{mLC} \\
m_{\rm IC}({\bf r})&=&-\frac{e}{\hbar c}\text{Im}\bra{{\bf r}}\mathcal{Q} x \mathcal{P} \mathcal{H} \mathcal{P} y \mathcal{Q}  \ket{{\bf r}},
\label{mIC} \\
m_{\rm BC}({\bf r})&=&\mu \frac{e}{2 \pi \hbar c}\mathfrak{C}(\bf r) . \label{mBC}  \label{mBC} \\
\mathfrak{C}({\bf r})&=&4\pi \text{Im}\bra{{\bf r}}\mathcal{Q} x \mathcal{P} y \mathcal{Q} \ket{{\bf r}}
\end{eqnarray}
It is worth noting that, despite this interpretation of the markers, only the macroscopic average of $m({\bf r})$ has physical significance, as discussed in Ref. \cite{Seleznev2023}. \textcolor{black}{The nomenclature of each contribution above follows the one established in the derivation based on the Wannier-function approach to the orbital-magnetization formula \cite{Thonhauser2005}. The term IC (itinerant current) is associated with the orbital magnetization contribution coming from the net current carried by Wannier functions that circulate through the surface of the solid. The LC term (local circulation) arises from the self-circulation of the Wannier functions. The BC (Berry curvature) term contains the explicit relation between orbital magnetization, Berry curvature, and the Chern number, as one can expect from equations (6) and (7), and makes contact with the Streda formula \cite{Bianco2013a, Streda1982}.}
In Ref. \cite{Wang2022}, it was shown that the orbital magnetization obtained from local markers coincides with that obtained from the definition of Eq.(\ref{Mdef}), provided the macroscopic average is taken over the entire sample area, including both bulk and edge regions. 
\begin{eqnarray}
M=\frac{1}{A}\int_{A}m({\bf r}) d{\bf r}\,\,.
\label{Mmarker}
\end{eqnarray}
This agreement holds even for small systems subjected to open boundary conditions (OBC). In each fractal generation, the integration area used in the average is given by $A=N_{\rm c}A_{\rm c}$, where the number of elementary cells in the structure, $N_c$, is half the number of sites  and $A_{\rm c}$ is the area of one cell. 
To calculate the orbital magnetization of fractal systems, we shall use both expressions [Eqs. (\ref{Mdef}) and (\ref{Mmarker})] to test this coincidence in the more intricate fractal geometry.

\section{Sierpinski Carpet} 
The Sierpinski carpet (SC) is geometrically constructed by systematically inserting vacancies inside a bidimensional region, creating different generations $G(n)$. By increasing the number of vacancies the SC dimension assumes a fractional value of $D=log(8)/log(3)\approx1.89$ independently of the generation. Adopting a honeycomb lattice mesh inside the carpet, the external lengths are fixed between the generations G(0$-$3) and the corresponding number of lattice sites are $N_i=11772, \ 10512, \ 9456, \ 8688$ sites $(i=0,...,3)$, respectively, as shown in gray, blue, red and green in Fig.~\ref{FIG1} (b). Both external and internal edges of the carpet are formed by sequences of armchair and zigzag contours. The resulting energy spectrum of SCs G(0$-$3) are exhibited in panel (c) of Fig.~\ref{FIG1} following the previously described colored pattern. The adopted parametrization was adjusted to match the topological phase ($C=+1$) inside the 2D Haldane's phase diagram with $V_{ab}/t_2=3.0$ and $\phi=0.7\pi$ [point (1) in Fig.~\ref{FIG1} (a)], which describes the anomalous quantum Hall effect. As expected from the bulk-edge correspondence \cite{Haldane1988}, in this phase, the finite system subjected to OBC has topologically protected edge states inside the bulk energy gap region [yellow shaded area]. Such states develop a staircase profile as the generation of the fractal is increased \cite{lage2025sc,Li2023}, which implies in emergent peaks in the total density of states (DOS) for this region, as shown in Fig.~\ref{FIG1} (c). In fractals this is characterized by emergent edge states occurring in external and internal edges of the system, the later being related with the fractal generation. For the proposed systems, this pattern results from a competition between the lattice parameters and the fractal pattern, emerging as the vacancies are inserted, as seen for different types of models and fractals \cite{lage2025sc}.

\begin{figure*}[t!]
    \centering
    \includegraphics[width=1.0\textwidth]{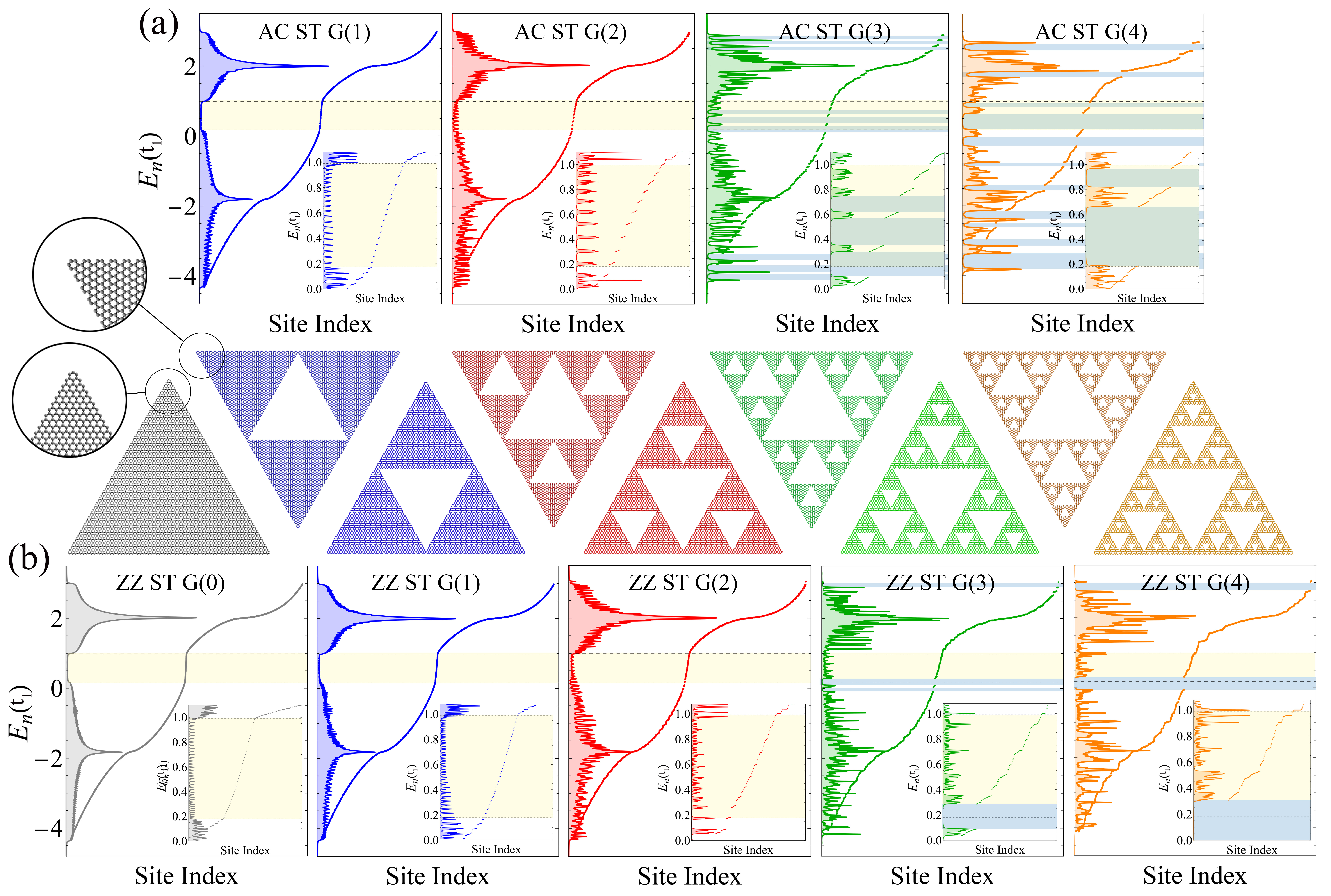} 
    \caption{\textcolor{black}{(a) Illustration of several generations of the considered Sierpinski Triangle with armchair edges [AC ST G(1$-$4)] and (b) with zigzag edges [ZZ ST G(0$-$4)]. Energy spectra with the corresponding density of states for AC and ZZ STs. The insets magnify the regions of in-gap states, which appear in the system subjected to open boundary conditions when parameters are adjusted to reach the topological phase of the model. The yellow shaded regions represent the band gaps of the 2D bulk Haldane model, obtained with the same coupling constants ($V_{ab}/t_2=3.0$ and $\phi=0.7\pi$) used in our calculations within the fractal geometry [point (1) in Fig.~\ref{FIG1} (a)]. The blue shaded strips correspond to the fractal induced energy gaps opened in the energy spectra of ZZ and AC STs.}}
   \label{FIG3}
\end{figure*}

In Fig.~\ref{FIG2+FIG4} (a.1$-$4) we present the calculated orbital magnetization, $M$, as a function of the chemical potential, $\mu$,  for the first four generations of the Sierpinski carpets, i.e. G(0$-$3). 
In each panel, the results of the two distinct computational methods are shown to be in perfect matching; the solid lines represent the magnetization calculated from its fundamental definition (Eq.~\ref{Mdef}), while the circles are obtained from the local marker with projectors formulation (Eqs. \ref{orbitalMarker} and \ref{Mmarker}). The shaded yellow region highlights the energy range of the bulk topological gap of the 2D Haldane model. For the initial G(0) generation in panel (a.1), the magnetization exhibits a linear trend within this gap, which is a signature of a finite bulk Chern number. It should be noted that the result in Fig.~\ref{FIG2+FIG4} (a.1), obtained here from the real-space calculation for a flake with OBC [SC G(0)], coincides with those obtained from the ${\bf k}$-space calculation in 2D bulk for the same set of parameters \cite{Ceresoli2006}. 
As the fractal generation increases from G(1) to G(3) significant fluctuations appears in panels (a.$2-4$) and become more frequent. This behavior is a direct consequence of the fractal nature of the system, reflecting the ``staircase'' profile of the several internal and external edge states that emerge in the energy spectrum, as highlighted in the magnified view of this energy region in the insets of Fig.~\ref{FIG2+FIG4} (a).




\begin{figure*}[t!]
\centering
\includegraphics[width=1.0\textwidth]{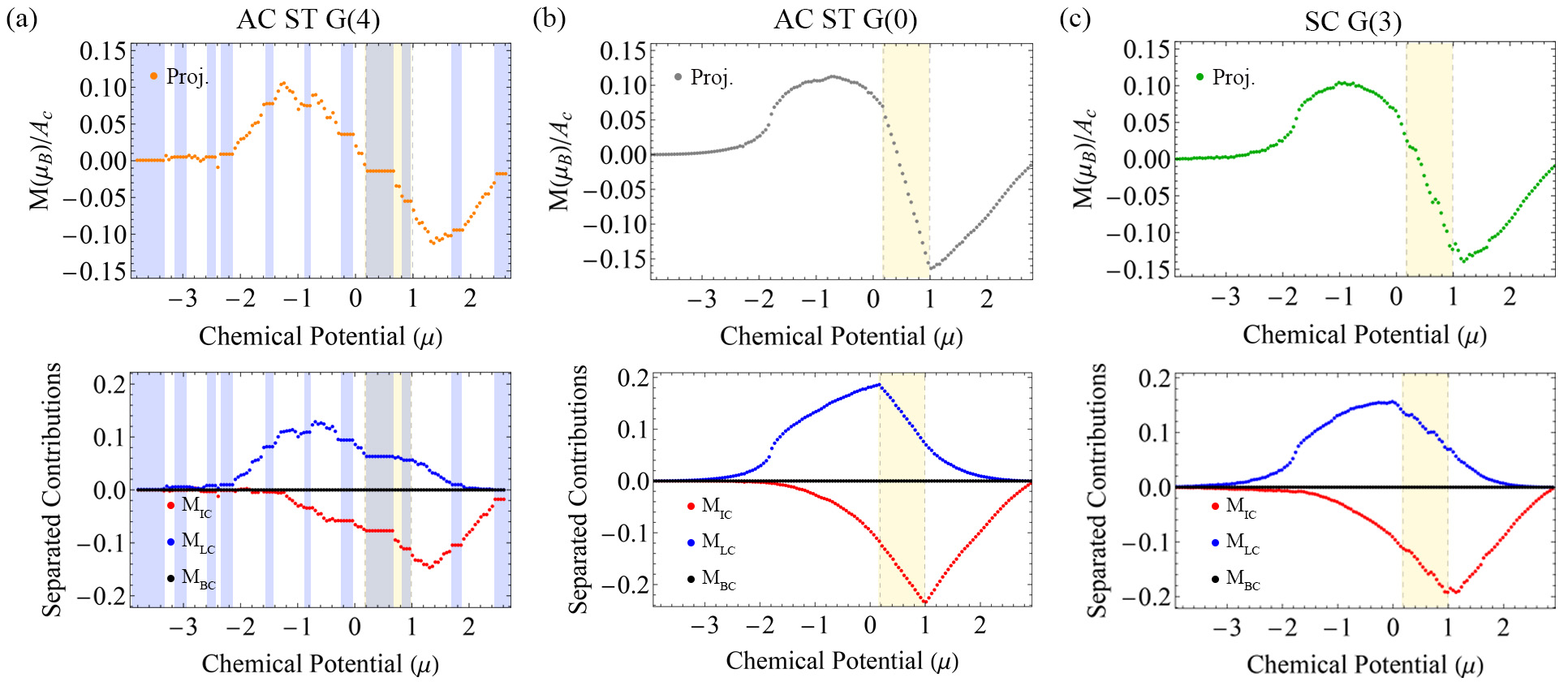}
\caption{Orbital magnetization calculated using local markers formulation (projectors formalism) for (a) ST AC G(4), (b) ST AC G(0) and (c) SC G(3) and its respective contributions for each component $M_{\text{IC}}$ (red dots), $M_{\text{LC}}$ (blue dots), and $M_{\text{BC}}$ (black dots). Haldane bulk gap and fractal induced gaps are highlighted, respectively, by regions in yellow and blue.}\label{nova7}
\end{figure*}

\section{Sierpinski Triangle} 

Similarly to the Sierpinski carpet, the Sierpinski triangle (ST) is constructed, exhibiting a fractional dimension of $D=\log(3)/\log(2) \approx 1.58$. Given the $C_3$ rotational symmetry of the triangular structures, by adopting a honeycomb lattice, we have two distinct edge configurations for the full Sierpinski triangle; zigzag (ZZ ST) and armchair (AC ST).  
As previously reported \cite{Pedersen2020}, the electronic properties of ZZ and AC STs are intrinsically distinct. The high sensitivity of ST quantum states, related to its edge configurations, comes directly from the fundamental physics of the quantum confinement in a honeycomb lattice.

{Fig.~\ref{FIG3} shows a comparison of the electronic spectrum of both AC and ZZ edge configurations in panels (a) and (b), respectively.  The number of sites in the AC and ZZ ST G(0$-$4) generations  are: $ N^{AC}_{0-4} = 7056, \ 5400,\ 4212,\ 3402,\ 2916  $ and $ N^{ZZ}_{0-4} = 6721, \ 5279,\ 4307,\ 3737,\ 3506  $, illustrated by gray, blue, red, green, and orange, respectively. \textcolor{black}{Since  both G(0) AC and ZZ STs have quite similar electronic spectra we have omitted the results regarding the AC ST G(0) energy spectra and DOS.} The parameters of the Haldane model are the same adopted for the SCs [point (1) in the phase diagram of Fig.~\ref{FIG1} (a)]. }

\textcolor{black}{Unlike SCs, fractality promotes additional gaps in the energy spectrum of STs. The presence of fractal gaps in STs and their absence in SCs are consistent with previous numerical works \cite{Pedersen2020,haldanecristiane,PhysRevB.93.115428}.} The principal fractal gaps in ST are highlighted by blue shaded areas along the energy states of Fig.~\ref{FIG3}. Within the yellow shaded region of the same results, a distinct staircase profile is also observed for some of the edge states residing inside the Haldane bulk gap. These states are separated by such fractal-induced gaps. For both types of ST, the fractal-induced gaps produce plateaus with constant orbital magnetization for different values of the chemical potential, as can be observed in Fig.~\ref{FIG2+FIG4} (b.3-4) and (c.3-4). Notably, some of these fractal gaps and orbital magnetization plateaus occur in an energetic region that overlaps with the Haldane bulk gap. In the next section, we further discuss this phenomenon.

\section{Discussion}

The more significant fractal-induced gaps and corresponding plateaus in orbital magnetization for both ZZ ST and AC ST are highlighted in blue shaded rectangles in Fig.~\ref{FIG2+FIG4} (b) and (c), respectively. Interestingly, for ST AC G(4), a large fractal gap overlaps with much of the 2D Haldane bulk gap, producing a wide plateau in the graph of orbital magnetization versus chemical potential. For both types of ST, the results obtained using the primary definition of orbital magnetization (Eq.~\ref{Mdef}), denoted by a continuous line, and those obtained from the local markers formulation (Eq. \ref{orbitalMarker} and \ref{Mmarker}), denoted by circles, coincide. \textcolor{black}{The features of orbital magnetization in fractal systems cannot be understood from the usual {\bf k}-space formulation, instead requiring a purely real-space perspective \cite{Wang2022}. In particular, an elegant interpretation of the orbital-magnetization plateaus in ST fractals can be obtained from the local-marker formulation presented in Eqs. (\ref{orbitalMarker}–\ref{Mmarker}). When the chemical potential $\mu$ passes through an energy range with no electronic states (fractal induced gaps marked in blue on real-space electronic spectra), the ground-state projector $\mathcal{P}$ and its complement $\mathcal{Q}$ remains invariant. Consequently, $m_{\rm LC}({\bf r})=\frac{e}{\hbar c}\text{Im}\langle{\bf r}\big|\mathcal{P} x \mathcal{Q} \mathcal{H} \mathcal{Q} y \mathcal{P}  \big|{\bf r}\rangle$, $m_{\rm IC}({\bf r})=-\frac{e}{\hbar c}\text{Im}\langle{\bf r}\big|\mathcal{Q} x \mathcal{P} \mathcal{H} \mathcal{P} y \mathcal{Q}  \big|{\bf r}\rangle$, and their corresponding sample average, $M_{\rm IC}=A^{-1}\int_{A}m_{\rm IC}({\bf r})d{\bf r}$ and $M_{\rm LC}=A^{-1}\int_{A}m_{\rm LC}({\bf r})d{\bf r}$, remain unchanged by $\mu$ within this energy range. The only term in Eq. (\ref{orbitalMarker}) that has an explicit dependence on $\mu$ is the term $m_{\rm BC}({\bf r})=\mu \frac{e}{2 \pi \hbar c}\mathfrak{C}(\bf r)$. Although it is locally modified by the chemical potential, the average over the entire sample must vanish, $M_{\rm BC}=A^{-1}\int_{A}m_{\rm BC}({\bf r})d{\bf r}=0$. This follows from the fact that the average of quantity $\mathfrak{C}({\bf r})$ vanishes when integrated over the whole system $A^{-1}\int_{A}\mathfrak{C}({\bf r})d{\bf r}=0$ [see the analysis in Appendix A], a well-known property of the Chern marker introduced by Bianco and Resta \cite{Bianco2011, Wang2022}. In Fig.~\ref{nova7} (a), we present the partial contribution of each term in Eq. (\ref{orbitalMarker}) to the total orbital magnetization [Eq. (\ref{Mmarker})] for the illustrative case of AC ST G(4) fractal systems. As explained above, the $M_{\rm BC}$ contribution vanishes at any $\mu$. In the Sierpinski triangles, the $M_{\rm LC}$ and $M_{\rm IC}$ terms display plateaus that reflect the total orbital-magnetization plateaus reported in previous results. The same idea holds for the magnetization plateau in the other ST fractals.}

\textcolor{black}{Fig.~\ref{nova7} (b) and (c) also present the total orbital magnetization and its separated contributions for the illustrative cases of AC ST G(0) and SC G(3) systems. In these cases, total magnetization also arises from the competition between the itinerant ($M_{\text{IC}}$) and local ($M_{\text{LC}}$) contributions. The Berry-curvature contribution vanishes in all cases because the local Chern marker, when integrated over the whole sample, also vanishes.}

These results illustrate the direct impact of the fractal geometry on the orbital magnetization. As the fractal generation increases, the orbital magnetization curve develops a more intricate structure, reflecting the emergence of a more complex hierarchy of energy structures in the electronic spectrum. 

\section{Conclusion} 
\textcolor{black}{A remarkable feature of STs geometry is the emergence of fractal-induced gaps \cite{Pedersen2020,haldanecristiane}.} Its iterative construction creates a self-organized hierarchy of boundaries, leading to quantum confinement that quantizes the electronic energy levels. \textcolor{black}{Such gaps have been shown to lead to interesting consequences for topology \cite{CristianePRL} and the formation of spin polarization \cite{Pedersen2020} in fractal systems. The orbital-magnetization plateaus found here add a new and noteworthy feature to these gaps.} The triangular geometry ensures each spatial side composed by a single, pure edge type either zigzag (ZZ) or armchair (AC), whose intrinsic properties dominate the electronic behavior. Consequently, switching the boundary type drastically modifies the energy spectra and orbital magnetization, which exhibit distinct plateaus within these fractal-induced gaps. In contrast to SCs fractals, which develop a dense, staircase-like spectrum and growing fluctuations in orbital magnetization, the STs emergent gaps promote constant orbital magnetization plateaus. This shows how quantum confinement in fractal geometry can be used to explore novel properties of electronic orbital angular momentum, offering an interesting perspective in the emerging field of orbitronics \cite{Cysne2025}. Proposals and experimental realizations of fractal geometry are available for different types of physical systems \cite{Chen2025, He2024, Li2017, Kempkes2018, Canyellas2024, Myerson-Jain2022}. Furthermore, the Haldane model, used here as a prototypical example, can be realized in different contexts, such as magneto-optical lattices \cite{Ablowitz2024}, ultracold fermion lattices \cite{Jotzu2014}, Fe-based insulators \cite{Kim2017}, and topoelectrical circuits \cite{Haenel2019}, opening a perspective to study the phenomenology discussed here.

\section*{Acknowledgments} The authors would like to thank the INCT de Nanomateriais de Carbono for providing support on the computational infrastructure. LLL thanks the CNPq scholarship. TPC acknowledges CNPq (Grant No. 305647/2024-5). AL thanks FAPERJ (Grant E-26/200.569/2023).

\appendix 
\setcounter{figure}{0}
\renewcommand{\thefigure}{A\arabic{figure}}

\section{Site-Resolved Local Chern Marker within Fractal-Induced Gaps}\label{A1}

\begin{figure}
    \includegraphics[width=8.7cm]{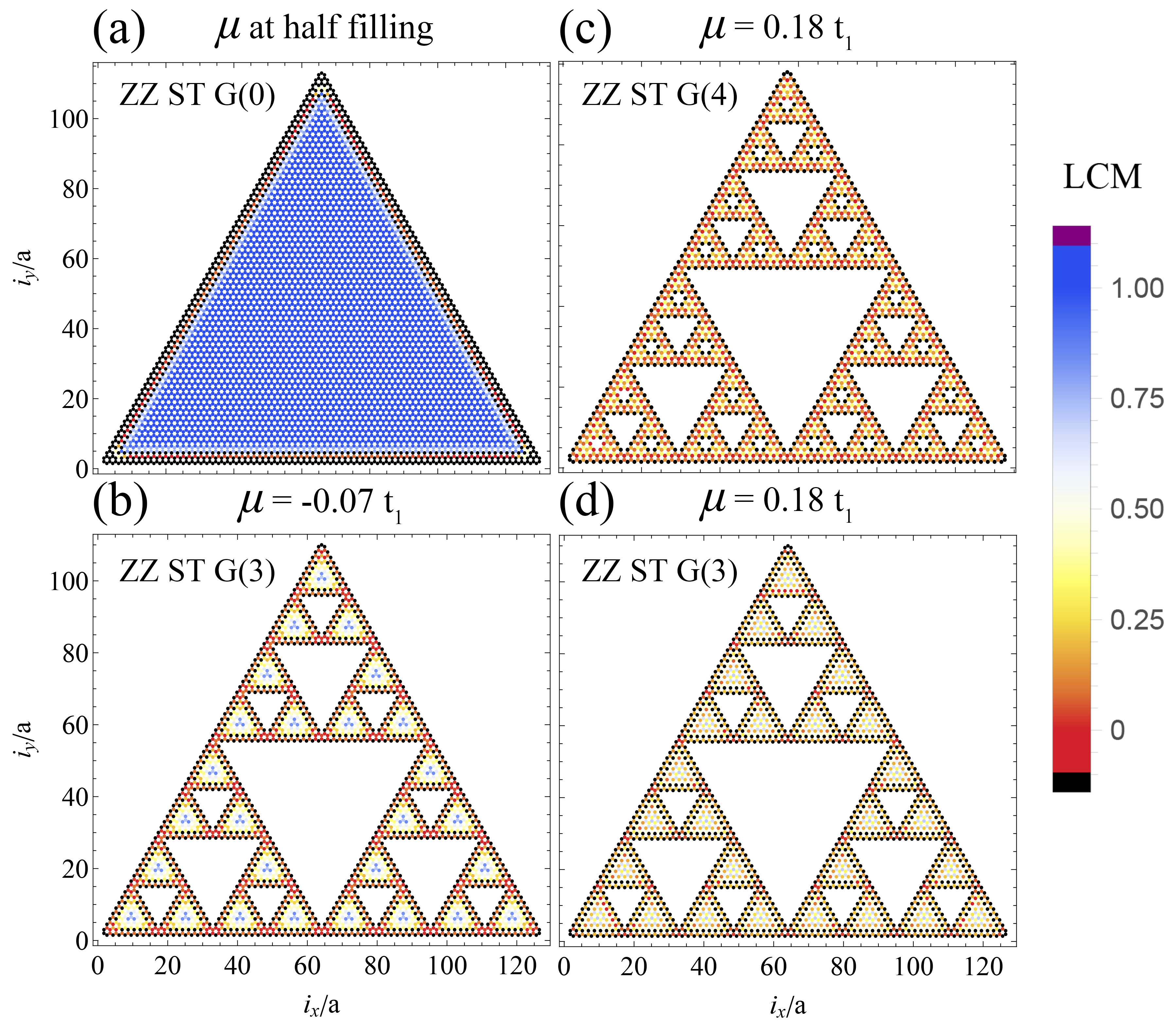} 
    \caption{Site-resolved local Chern marker for (a) ZZ ST G(0) at half filing, (b) ZZ ST G(3) with $\mu = -0.07t_1$, (c) ZZ ST G(3) with $\mu = +0.18t_1$ and (d) ZZ ST G(4) with $\mu = +0.18t_1$. Note that both negative and positive values, marked as black and purple sites, respectively, are present in the site-resolved map. The legend color bar is cut at the range, which is between 0 and 1 values. The negative values of the marker compensate for the positive values in such a way that the whole-sample average of $\mathfrak{C}({\bf r})$ vanishes.}
   \label{FigA1}
\end{figure}

\textcolor{black}{Differently from the ${\bf k}$-space formula of orbital magnetization, valid for uniform bulk system, where plateaus are associated with a null Chern number \cite{Thonhauser2005}, for the Sierpinski triangles the plateaus in orbital magnetization can coexist with a non-null Local Chern Marker as can be understood with a real space projection.} Fig.~\ref{FigA1} shows the site-resolved local Chern marker (LCM) calculated for ST considering distinct fractal generations and chemical potentials. In panel (a) of Fig.~\ref{FigA1}, we show ZZ ST G(0) at half filling. In this case, the center of the sample has a quantized positive value of the Chern marker, whereas at the edges, the Chern marker assumes negative values. This profile is expected for the local Chern marker for a uniform system subjected to OBC in the Chern insulator phase. The negative values of the marker at the edges compensate for the positive value in the middle of the sample in such a way that the whole-sample average of the marker vanishes, as discussed in Ref. \cite{Bianco2011}. In panels (b), (c), and (d) of Fig.~\ref{FigA1}, we show the site-resolved local markers for higher generations of ZZ ST with the chemical potential in the middle of the fractal induced gaps, i.e., (b) ZZ ST G(3) with $\mu = -0.07t_1$, (c) ZZ ST G(4) with $\mu = +0.18t_1$, and (d) ZZ ST G(3) with $\mu = +0.18t_1$. For all three cases, the local Chern marker assumes finite values at the sites, but the whole-sample average of the marker also vanishes.

\setcounter{figure}{0}
\renewcommand{\thefigure}{B\arabic{figure}}

\section{Additional results on orbital magnetization in ST}\label{A2}

\begin{figure}[h!]
\centering
\includegraphics[width=8.7cm]{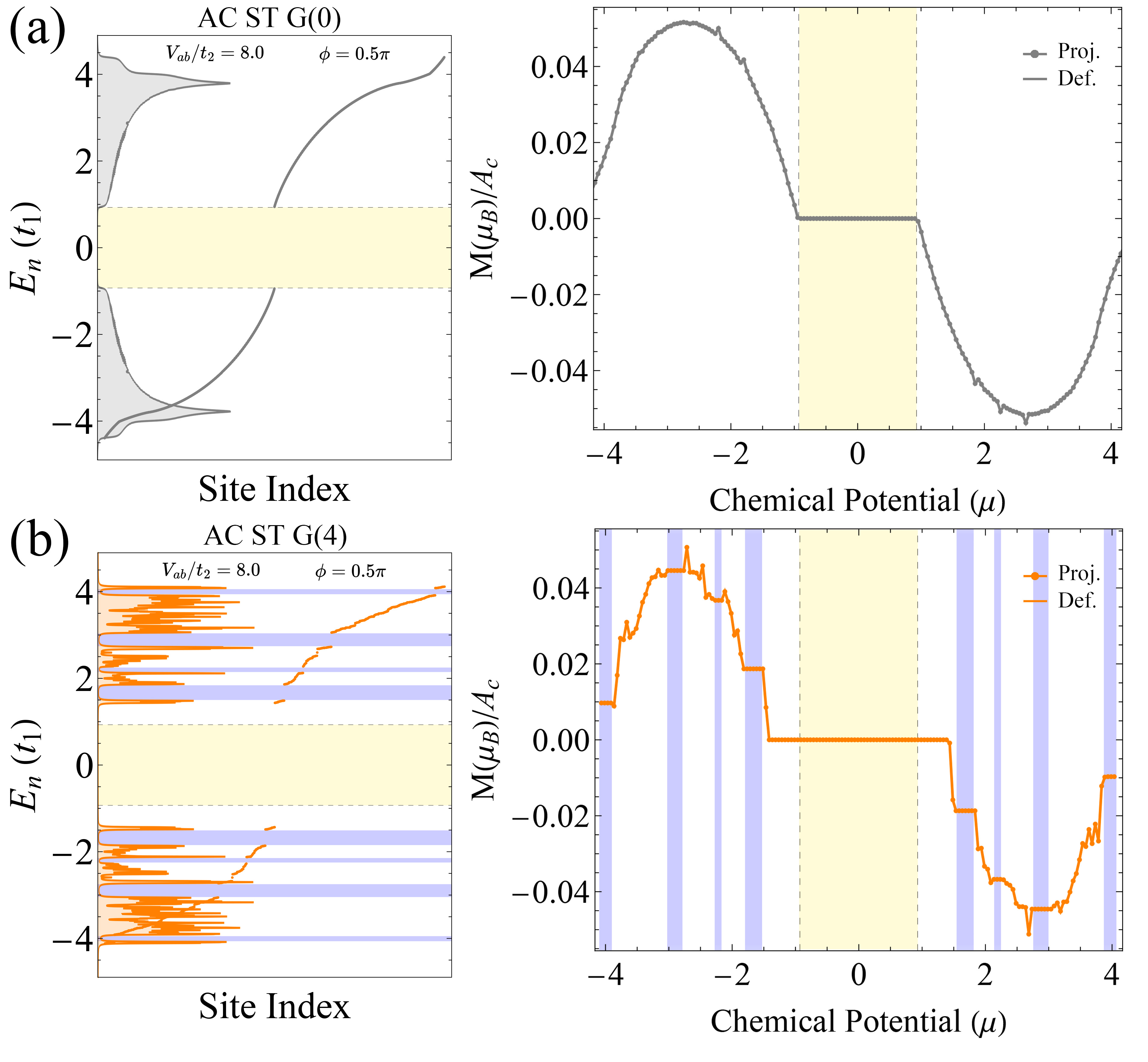}
\caption{Energy spectrum and orbital magnetization for model parameters set at point (2) in the trivial phase of the topological phase diagram shown in Fig. 1(a): $t_2=t_1/3$, $V_{ab}/t_2=8.0$ and $\phi=0.5\pi$. The results are shown for AC ST (a) G(0) and (b) G(4). The Haldane bulk gap is highlighted in yellow, and the fractal induced gaps in blue.
}\label{nova6}
\end{figure}

\textcolor{black}{Here we extend our analysis by exploring a trivial point in the bulk Haldane phase diagram shown in Fig.~\ref{FIG1} (a). In the 2D (bulk) system, the Haldane model presents particle-hole symmetry for $\phi = 0.5\pi$. Then we consider this value of $\phi$ and a representative value of the sublattice potential $V_{ab}/t_2 = 8.0$, which drives the system to the topologically trivial phase, marked as point (2) in Fig.~\ref{FIG1} (a). In Fig.~\ref{nova6} we present the electronic spectra and the orbital magnetization as a function of $\mu$ for the AC ST G(0) and G(4) in panels (a) and (b), respectively. 
The overall phenomenology follows the one detailed in the main text, except that in the trivial-insulator case, the states invading the Haldane bulk-gap region (yellow area) are absent in the finite-system energy spectra. In addition, for the parameters shown in Fig.~\ref{nova6}, fractality slightly enlarges the central gap. In the bulk-state energy bands, fractal-induced gaps characterized by orbital magnetization plateaus also occur, being represented in the blue regions. A similar behavior is also observed in ZZ ST in the topologically trivial phase and is therefore not presented here to avoid repetition.}

\begin{figure*}
\centering
\includegraphics[width=180mm]{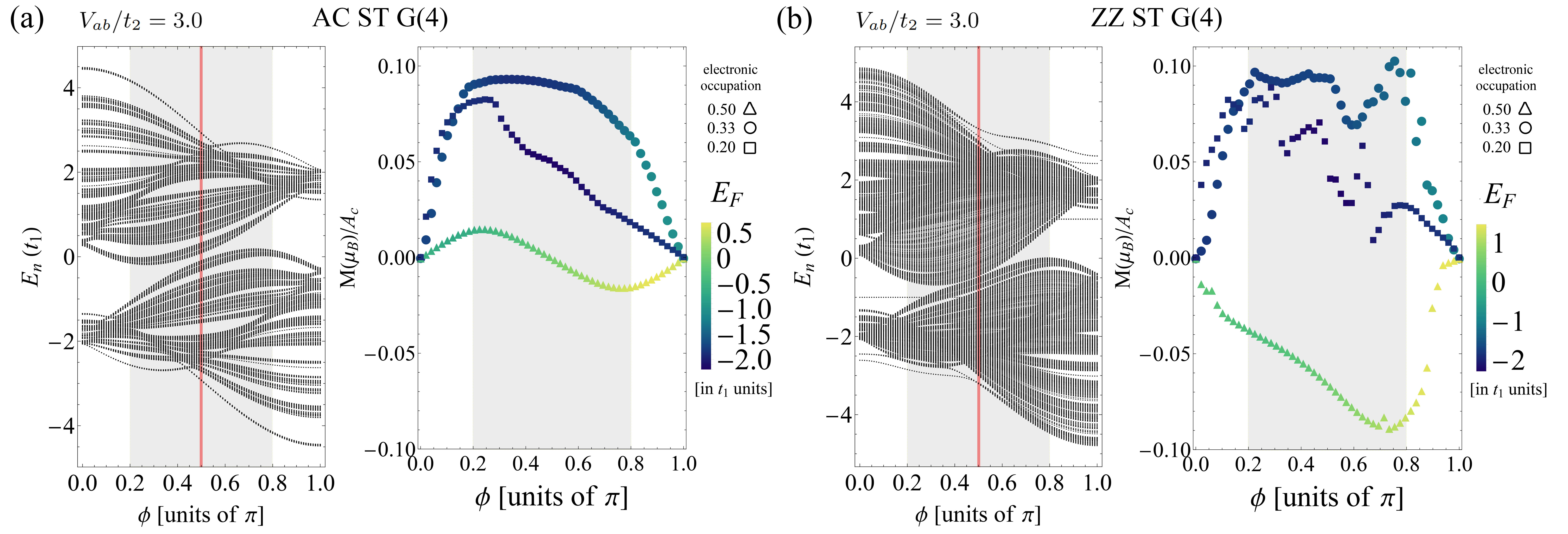}
\caption{
Energy spectrum and orbital magnetization of (a) AC and (b) ZZ ST G(4) lattices as a function of   $\phi$. In both cases $t_2=1/3$ and $V_{ab}/t_2=3.0$. The vertical red lines in the plots of the spectra indicate the value $\phi=\pi/2$. In the plots of magnetization normalized by the area $A_c$, three distinct electronic occupations are considered: $\nu=0.50$ (triangles), $0.33$ (circles) and $0.20$ (squares).  The color scale indicates the Fermi level $E_F$ (in units of $t_1$) for each occupation. The shaded gray areas designate the Chern insulator regions in the 2D Haldane phase diagram. The remaining portions in white corresponds to a trivial insulator.
}\label{mag_eig_vs_phi}
\end{figure*}

\textcolor{black}{In Fig.~\ref{mag_eig_vs_phi} we present the evolution of the electronic spectra and the orbital magnetization at fixed electronic occupations $\nu$ (0.20,0.33,0.50) as the phase $\phi$ of the Haldane model is varied. The results are shown for two illustrative cases, the AC ST G(4) and ZZ ST G(4) with $t_2=t_1/3$ and $V_{ab}/t_2=3.0$. Notably, for triangular systems with AC edges, the number of sites on sublattices A and B is equal. Nevertheless, in triangular systems with ZZ edges, there is an excess of sites on one of the sublattices. For this reason, the electronic spectra of AC ST G(4) preserve particle–hole symmetry when $\phi=\pi/2$, whereas this symmetry is lost for ZZ ST G(4) with finite $V_{ab}$. This effect arising from an imbalance in the number of sites in different sublattices and its signatures in the electronic spectra of fractals are discussed in detail in Ref. \cite{haldanecristiane}. This distinction between fractals with different terminations is reflected in the contrasting magnetization curves as a function of $\phi$ for similar electronic occupations, as reported in Fig.~\ref{mag_eig_vs_phi}.}

\bibliography{refs}

\end{document}